\documentclass[12pt]{iopart}

\usepackage{graphics}
\begin{document}

\article{COMMENT}{Comment on 'Excitation function for the radionuclide $^{46}$Sc produced in the irradiation of $^{45}$Sc with deuterons and $^6$He'}

\author{M~Avrigeanu, V~Avrigeanu}

\address{"Horia Hulubei" National Institute for Physics and Nuclear Engineering, P.O. Box MG-6, 077125 Bucharest-Magurele, Romania}

\ead{\mailto{mavrig@nipne.ro}}

\begin{abstract}
We show that the model analysis of new measured $(d,p)$, $(d,t)$, ($^6$He,$^5$He), and ($^6$He,$^4$He) reaction cross sections at incident energies around the Coulomb barrier ({\bf {\it J. Phys. G: Nucl. Part. Phys.} 38 (2011) 035106}) led to results that are not consistent with similar calculated and evaluated data. On the other hand, it should be corrected by taking into account the direct processes.
\end{abstract}
\submitto{\JPG}
\maketitle

The cross sections of the $^{45}$Sc$(d,p)$$^{46}$Sc and $^{45}$Sc$(d,t)$$^{44}$Sc reactions have recently been measured at incident energies around the Coulomb barrier, namely between 0.9 and 11.7 MeV, and compared with results of pre-equilibrium and statistical model calculations by Skobelev et al. \cite{skobelev11}. While the measurements at deuteron energies close to the Coulomb barrier and the comparative discussion of the $(d,p)$ and $(d,t)$ one-nucleon transfer reactions are of large interest, the model calculations described in this paper rise several question marks.

Firstly, the pre-equilibrium emission (PE) and compound nucleus (CN) processes were considered by using the well known codes TALYS \cite{TALYS} and EMPIRE \cite{EMPIRE}. However, for incident energies below and around the Coulomb barrier, the deuteron interaction with target nuclei proceeds largely through direct reaction processes while the PE and CN reaction mechanisms start to prevail at larger incident energies. This has recently been shown through an unitary and consistent quantitative description of deuteron complex interactions with $^{27}$Al, $^{63,65}$Cu and $^{93}$Nb \cite{ma09,bem09,ma10a,ma10b,es10}, for all reaction cross sections measured for a given target nucleus, including the $(d,p)$ and $(d,n)$ reactions. Only to verify this statement in the case of $^{45}$Sc nucleus too, the new measured data \cite{skobelev11} are compared in Fig. 1 with the results obtained with TALYS-1.2 for the $^{45}$Sc(d,p)$^{46}$Sc and $^{45}$Sc(d,t)$^{44}$Sc excitation functions. We have used in this respect the whole TALYS default input parameter set and, additionally, the adjusted value $r_{vadjust}$=1.12 adopted in Ref. \cite{skobelev11} for the reduced radius of the optical model (OM) volume potentials. The adjustment has been applied to both deuteron and proton OM potentials since no definite notice was given by Skobelev et al., the corresponding effects leading together to an increase of the $(d,p)$ reaction cross section from $\sim$40\% around the Coulomb barrier to less than 20\% at 11.7 MeV. However, even these larger values underestimate the experimental data by a factor ranging from 2 at the lower energies to more than 4 at 11.7 MeV. Especially above the Coulomb barrier they are rather close to the content of TENDL 2010 library \cite{TENDL}, obtained by using the TALYS code and released in December 2010. Conversely, these evaluated data as well as those we have calculated are smaller by a factor $\sim$2 with respect to the similar curve shown in Fig. 2 of Ref. \cite{skobelev11}. As a result, this curve could not be replicated. 

\begin{figure}  [t]
\resizebox{0.8\textwidth}{!}{\includegraphics{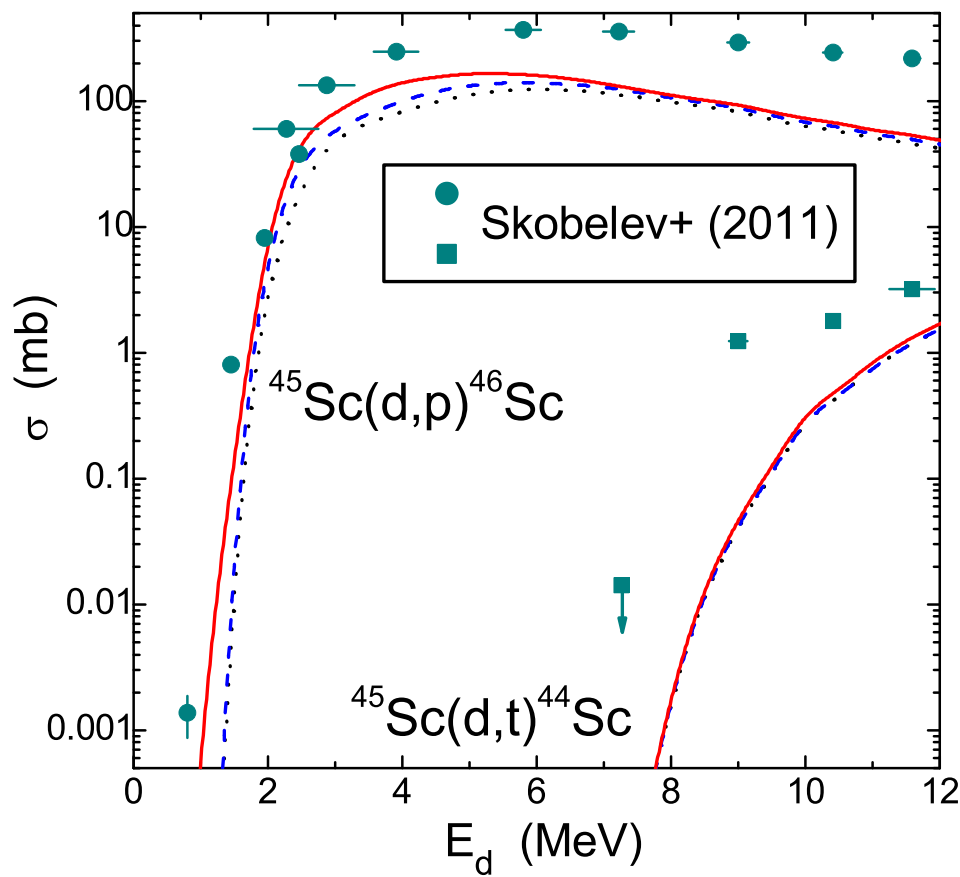}}
\caption{\label{Fig1}(Color online) Comparison of measured excitation functions for the reactions $^{45}$Sc(d,p)$^{46}$Sc (dots) and  $^{45}$Sc(d,t)$^{44}$Sc (triangles) \cite{skobelev11}, and the calculated results of the code TALYS-1.2 \cite{TALYS} with default input parameters (dotted curves) as well as adjusted \cite{skobelev11} OM potential volume radius for deuterons (dashed) and deuterons and protons (solid).}
\end{figure}

On the other hand, the condition of a consistent input parameter set could be fulfilled only if the adopted change is validated by an OM analysis of elastic scattering data (e.g., \cite{ma09,bem09}) or at least of total reaction cross sections $\sigma_R$ below the Coulomb barrier. However, this has not been the case of the above-mentioned OM reduced-radius adjustment by Skobelev et al. Moreover, a systematic analysis of the proton $\sigma_R$ in the same mass region showed that the default proton OM potential of TALYS does not underestimate this quantity but eventually overestimates it \cite{ma08a}.

As a matter of fact the underestimation of the $^{45}$Sc(d,p)$^{46}$S experimental data by PE and CN calculations, using the TALYS or EMPIRE codes, could be really expected due to the related absence of a proper consideration of the direct stripping mechanism. This could be obtained using, e.g., the Coupled Reaction Channels formalism \cite{FRESCO}. This way is experimentally endorsed by population of more than 80 discrete levels up to $\sim$4 MeV excitation energy in $^{46}$Sc \cite{levs46sc} by 7 MeV deuterons \cite{rapaport66}, as well as of $\sim$200 discrete levels up to $\sim$7 MeV excitation energy by 12 MeV deuterons \cite{roy92}. Therefore the strong direct stripping mechanism contribution is probably hidden by PE and CN parameters alteration that should have been considered by Skobelev et al. in Ref. \cite{skobelev11}.

We show in Fig. 1 also the results of a similar analysis of the $^{45}$Sc(d,t)$^{44}$Sc reaction, at energies where the model calculation sensitivity to OM potential radius is much lower. These results are smaller by a factor of at least 2 than the measured data but rather identical with the TENDL-2010 data. However, the underestimation of the $(d,t)$ data by TALYS or EMPIRE calculations is the result of the direct process type of this channel, which is proved by spectroscopic studies of the $^{44}$Sc discrete levels strongly populated through $(d,t)$ pick-up \cite{levs44sc}. Actually, the lowest energy side of a $(d,t)$ excitation function, between its threshold and those of the $(d,p2n)$ and $(d,dn)$, can be described exclusively within the pick-up reaction mechanism as it is shown in Fig. 3 of Ref. \cite{ND2010}. Nevertheless, the additional consideration of this reaction is consistent with an unitary analysis of nuclear model predictions taking into account all available data for various reaction channels (e.g., \cite{bem09,ma10a,ND2010,ma05,maCDCC10}). On the other hand, the evaluated \cite{TENDL} and presently calculated data are larger by a factor of $\sim$2 with respect to the TALYS results of Skobelev {\it et al.}. Therefore, we could not replicate these results either. The latest additional reduction of their $(d,t)$ cross sections values might be a consequence of the stronger enhancement of $(d,p)$ reaction channel in Ref. \cite{skobelev11}. 

Finally, the analysis of the $^{45}$Sc($^6$He,$^5$He)$^{46}$Sc and $^{45}$Sc($^6$He,$\alpha)$)$^{47}$Sc excitation functions were carried out by Skobelev {\it et al.} using the EMPIRE code and the mentioned circumstance of the only CN mechanism consideration (Sec. 3 of Ref. \cite{skobelev11}) in spite of their own note in Sec. 4 on 'the predominant contributions of the direct reactions'. Moreover, there is no comment on the large data underestimation generated by these model calculations. However, it is well known that the halo character of $^6$He Borromean projectile makes it a particularly interesting object of both structure and reaction model studies \cite{keeley09} so that the new data \cite{skobelev11} may be used to check the $^6$He structure models as well as to comparatively analyse the sequential transfer via the($^6$He,$^5$He; $^5$He,$^4$He) process, the one-step transfer of a di-neutron pair ($^6$He,$^4$He), and the $\alpha$-particle evaporation \cite{keeley09,alamanos}. Therefore, the nuclear model analysis of the valuable new measured $(d,p)$, $(d,t)$, ($^6$He,$^5$He), and ($^6$He,$^4$He) reaction data \cite{skobelev11} should definitely include the direct processes consideration that is crucial at incident energies around Coulomb barrier.

\section*{Acknowledgments}
This work was supported by the CNCSIS project No. PNII-IDEI-43/2008.


\section*{References}
\begin{thebibliography}{99} 

\bibitem{skobelev11} Skobelev N K, Kulko A A, Kroha V, Burjan V, Hons Z, Daniel A V, Demeskhina N A, Kalpakchieva R, Kugler A, Mrazek J, Penoinzhkevich Yu E, Piskor S, Simeckova E and  Voskoboynik E I 2011 J. Phys. G: Nucl. Part. Phys. {\bf 38} 035106
\bibitem{TALYS} Koning A J, Hilaire S and Duijvestijn M C  2008 TALYS-1.0  {\it Proc. Int. Conf. on Nuclear Data for Science and Technology, Nice, 2007} ed. O Bersillon {\it et al.} (EDP Sciences, Paris)  211; version TALYS-1.2, December 2009; http://www.talys.eu/home/ 
\bibitem{EMPIRE} Herman M, Capote R, Carlson B V, Oblozinsky P, Sin M, Trkov A, Wiencke H and Zerkin V 2007 Nucl. Data Sheets {\bf 108} 2655
\bibitem{ma09} Avrigeanu M, von Oertzen W, Forrest R A, Obreja A C, Roman F L and Avrigeanu V 2009 Fusion Eng. Design {\bf 84} 418
\bibitem{bem09} B\'em P, \v Sime\v ckov\' a E, Honusek M, Fischer U, Simakov S P, Forrest R A, Avrigeanu M, Obreja A C, Roman F L and Avrigeanu V 2009 Phys. Rev. C {\bf 79}  044610
\bibitem{ma10a} Avrigeanu M and Avrigeanu V 2010 EPJ Web of Conferences {\bf 2} 01004 
\bibitem{ma10b} Avrigeanu M and Avrigeanu V 2010 J. Phys: Conf. Ser. {\bf 205} 012014 
\bibitem{es10} \v Sime\v ckov\' a E, B\'em P, G\"otz M, Honusek M, Mr\'azek J, Nov\' ak J, \v Stefánik M, Z\' avorka L, Avrigeanu M and Avrigeanu V 2010 EPJ Web of Conferences {\bf 8} 07002 
\bibitem{TENDL} Koning A J and Rochman D 2010 TENDL-2010: TALYS-Based Evaluated Nuclear Data Library (December 8, 2010), http://www.talys.eu/tendl-2010/
\bibitem{ma08a} Avrigeanu M 
et al. (2008) Nucl. Phys. A {\bf 806} 15
\bibitem{FRESCO} Thompson I J 1988 Comput. Phys. Rep. {\bf 7}  167 (1988); Version FRES 2.3, May 2007 
\bibitem{levs46sc} S -C Wu 2000 Nucl. Data Sheets {\bf 91} 1
\bibitem{rapaport66} Rapaport R, Sperduto A and Bruechner W W 1966 Phys. Rev. {\bf 151}  939 
\bibitem{roy92} Roy J N, Majunder A R and Sen Gupta H M  1992  Phys. Rev. C {\bf 46}  144
\bibitem{levs44sc} Cameron J A and Balraj Singh 1999 Nucl. Data Sheets {\bf 88}  299 
\bibitem{ND2010} Avrigeanu M and Avrigeanu V 2011 Proc. Int. Conf. on Nuclear Data for Science and Technology, Jeju, Koreea, 2010 (in press)
\bibitem{ma05} Avrigeanu M, von~Oertzen W, Fischer U and Avrigeanu V 2005 Nucl. Phys. A {\bf 759} 327 
\bibitem{maCDCC10} Avrigeanu M and Moro A M 2010 Phys. Rev. C {\bf 82} 054605 
\bibitem{keeley09} Keeley N, Alamanos N, Kemper K W and Rusek R 2009 Prog. Part. Nucl. Phys {\bf 63}  396 
\bibitem{alamanos} Keeley N, Alamanos N and Lapoux V 2004 Phys. Rev. C {\bf 69}  064604 
\end{thebibliography}
\end{document}